\documentclass[12pt]{article}
\usepackage[english]{babel}
\usepackage{latexsym}
\usepackage{amsfonts}
\begin{document}

\begin{center}
{\Large \bf Brane-induced gravity in warped backgrounds\\ and the
absence of the radion}\\

\vspace{4mm}

Mikhail N.~Smolyakov\\

\vspace{4mm}

Physics Department, Moscow State University,\\ Vorob'evy Gory,
119992 Moscow, Russia
\\
\end{center}

\begin{abstract}
The Randall-Sundrum model with brane-localized curvature terms is
considered. It is shown that this model possesses some interesting
features, in particular, the radion field is absent in it.
Although there is no modification of gravity at long distances,
the model predicts deviations from Newton's law at short
distances. This effect can be observed in the experiments for
testing  gravity at sub-millimeter scales. \vspace{0.3cm}
\end{abstract}

\section{Introduction}
Models with brane-localized curvature terms have been widely
discussed in the literature during the last few years. In
paper \cite{DGP} it was argued that matter on the brane can
induce a brane-localized curvature term via the quantum corrections,
which appears in the low-energy effective action. An interesting
feature of this model is a modification of gravity at
ultra-large scales, which can be very interesting from the
cosmological point of view. But later it was shown
\cite{Luty,Rubakov} that  there exists a strong
coupling effect in this model, which makes it unacceptable. There were
attempts to merge the DGP-proposal and models with
warped backgrounds to get a long-distance modification of gravity.
But it turnes out that such models must be rejected for some reasons
(see, for example, \cite{Luty,DubLib,Padilla}). For example, the
Randall-Sundrum model with brane-localized curvature terms admits
a long-distance modification of gravity, but in this case either the
radion, the graviton or both fields become ghosts.

It would be interesting to consider such models from another point
of view. A modification of gravity at large distances is not the
only interesting effect, which can arise in the models. For
example, in  papers \cite{KTT,DHR} the spectrum of Kaluza-Klein
gravitons in the RS background with brane-localized terms for
different values of parameters was studied and some experimental
constraints were found (for example, for  collider
experiments), but the radion field was not taken into account. As
it was noted above, this field plays an important role in the
spectrum of gravitational fluctuations and  its existence
can change some parameters of the model considerably to make it
acceptable from the phenomenological point of view. One can recall
the original Randall-Sundrum model with two branes \cite{RS1}, in
which it is necessary to stabilize the size of extra dimension and
to make the radion field massive (see \cite{BKSV,SV}), for example,
with the help of the Goldberger-Wise mechanism \cite{wise}.

In the present  paper we study a model with brane-localized
curvature terms, which is based on the Randall-Sundrum solution
for the background metric. We will show that, with  appropriate
parameters, the model reproduces 4-dimensional gravity on the
brane, does not contradict the known experimental data and
provides some interesting consequences.

\section{The setup}
Let us choose the action of the model in the following form
\begin{equation}\label{actionRS}
 S = S_g + S_1 + S_2,
\end{equation}
where $S_g$, $S_1$ and $S_2$ are given by
\begin{eqnarray}\label{actionsRS}
S_g&=& \frac{1}{16 \pi \hat G} \int_E
\left(R-\Lambda\right)\sqrt{-g}\, d^{4}x dy,\\ \nonumber
 S_1&=&  \frac{\alpha_{1}}{16 \pi \hat G}\int_E \sqrt{-\tilde g}(\tilde R- \Lambda_1)
\delta(y) d^{4}x dy,\\ \nonumber
 S_2&=&  \frac{\alpha_{2}}{16 \pi \hat G}\int_E \sqrt{-\tilde g}(\tilde R-\Lambda_2)
\delta(y-R) d^{4}x dy.
\end{eqnarray}
{Here $\tilde g_{\mu\nu}$ is the induced metric on the branes and
the subscripts 1 and 2 label the branes.} The model possesses the
usual $Z_{2}$ orbifold symmetry. We also note that the signature
of the metric $g_{MN}$ is chosen to be $(-,+,+,+,+)$. Obviously,
the model admits the Randall-Sundrum solution for the metric,
which has the form
\begin{equation}\label{metricrs}
ds^2=  \gamma_{MN} d{x}^M d{x}^N = \gamma_{\mu\nu} {dx^\mu dx^\nu} +  dy^2,
\end{equation}
where $\gamma_{\mu\nu}=e^{2\sigma(y)}\eta_{\mu\nu}$,
$\eta_{\mu\nu}$ is the Minkowski metric and {the function}
$\sigma(y) = -k|y|$ in the interval $-R \leq y \leq R$. The
parameter  $k$ is positive and has the dimension of mass, the
parameters $\Lambda$ and $ \Lambda_{1,2}$, ${\alpha_{1,2}}$ are {related to it as
follows:}
\begin{equation}\label{Lambda}
 \Lambda = -k\alpha_{1}\Lambda_1= k\alpha_{2}\Lambda_2 = -12 k^2.
\end{equation}
 The function $\sigma$ has the properties
\begin{equation}\label{sigma}
  \partial_4 \sigma = -k\, sign(y), \quad \partial^2_{4}\sigma
  =-2k(\delta(y) - \delta(y-R)) \equiv  -2k\tilde\delta .
\end{equation}

The parameters ${\alpha_{1}}$ and ${\alpha_{2}}$  are not specified by the solution,
and their possible values will be duscussed below.

We denote $\hat \kappa = \sqrt{16 \pi \hat G}$, where $\hat G$ is
the five-dimensional gravitational constant, and parameterize the
metric $g_{MN}$ as
\begin{equation}\label{metricpar}
  g_{MN} = \gamma_{MN} + \hat \kappa h_{MN},
\end{equation}
$h_{MN}$ being the metric fluctuations. In  papers \cite{BKSV,SV} the second
variation Lagrangian for the fluctuations of metric in the
Randall-Sundrum model was obtained. In the case under consideration the presence of
the brane-localized curvature terms changes this Lagrangian, and the addition can be easily
calculated. But even with this addition the corresponding action is invariant under the
gauge transformations
\begin{eqnarray}\label{gaugetrRS}
h'_{MN}(x,y) = h_{MN}(x,y) -(\nabla_M\xi_N(x,y) +
\nabla_N\xi_M(x,y) ),
\end{eqnarray}
where $\nabla_M$ is the covariant derivative with respect to the
background metric $\gamma_{MN}$, and the functions $\xi_N(x,y)$
satisfy the orbifold symmetry conditions
\begin{eqnarray}\label{orbifoldsym1}
\xi^{\mu}\left(x,-y\right)&=&\xi^{\mu}\left(x,y\right),\\
\nonumber \xi^{4}\left(x,-y\right)&=&-\xi^{4}\left(x,y\right).
\nonumber
\end{eqnarray}
With the help of these gauge transformations we can impose the
gauge
\begin{equation}\label{unitgauge}
h_{\mu4} =0, \, h_{44} = h_{44}(x) \equiv \phi (x),
\end{equation}
which  will be called the {\it unitary gauge} (see \cite{BKSV}).
We would like to emphasize once again that the branes remain
straight in this gauge, i.e. we {\it do not} use the bent-brane
formulation, which allegedly destroys the structure of the model
(this problem was discussed in \cite{AIMVV}).

First,  let us consider the case, where there is no  matter
on the branes. In this case the equations of motion  for different
components of the metric fluctuations in the unitary gauge take
the form:

 1) $\mu\nu$-component
\begin{eqnarray}\label{mu-nu}
 & &\frac{1}{2}\left(\partial_\rho \partial^\rho h_{\mu\nu}-
\partial_\mu \partial^\rho
h_{\rho\nu}-\partial_\nu \partial^\rho h_{\rho\mu} +
\partial^2_{4} h_{\mu\nu}\right)- \\ \nonumber
&-& 2k^2h_{\mu\nu}+\frac{1}{2}\partial_\mu
\partial_\nu\tilde h+ \frac{1}{2}\partial_\mu \partial_\nu \phi+
\\ \nonumber &+& \frac{1}{2} \gamma_{\mu\nu}\left(\partial^\rho
\partial^\sigma h_{\rho\sigma}-\partial_\rho \partial^\rho \tilde
h - \partial^2_{4} \tilde h-4\partial_4 \sigma\partial_4 \tilde h
- \partial_\rho \partial^\rho \phi + 12 k^2 \phi\right)+\\
\nonumber &+& \left[2k  h_{\mu\nu} - 3k\gamma_{\mu\nu}\phi
\right]\tilde \delta +\\ \nonumber &+&
\frac{\alpha_{i}}{2}\delta_{i}\left[\left(\partial_\rho
\partial^\rho h_{\mu\nu}-
\partial_\mu \partial^\rho
h_{\rho\nu}-\partial_\nu \partial^\rho h_{\rho\mu}+\partial_\mu
\partial_\nu\tilde h\right)\right.+\\ \nonumber &+&\left.\gamma_{\mu\nu}\left(\partial^\rho
\partial^\sigma h_{\rho\sigma}-\partial_\rho \partial^\rho \tilde
h\right)\right] = 0,
\end{eqnarray}
where $i=1,2$, $\delta_{1}=\delta(y)$, $\delta_{2}=\delta(y-R)$

 2) $\mu 4$-component,
\begin{equation}\label{mu-4}
\partial_4 ( \partial_\mu \tilde h - \partial^\nu  h_{\mu\nu})-
3\partial_4 \sigma \partial_\mu \phi = 0,
\end{equation}
which plays the role of a constraint,

 3) $4 4$-component
\begin{equation}\label{4-4}
\frac{1}{2}(\partial^\mu \partial^\nu  h_{\mu\nu} - \partial_\mu
\partial^\mu \tilde h ) - \frac{3}{2}\partial_4 \sigma \partial_4 \tilde h
+ 6 k^2 \phi =0,
\end{equation}
where $\tilde h=\gamma^{\mu\nu}h_{\mu\nu}$. In what follows, we
will also use an auxiliary equation, which is obtained by
multiplying the equation for $44$-component by 2 and subtracting
it from the contracted equation for $\mu\nu$-component. This
equation contains $\tilde h$ and $\phi$ only and has the form:
\begin{eqnarray}\label{contracted-44}
& &\partial^2_{4} \tilde h + 2\partial_4 \sigma \partial_4 \tilde
h -8k^2 \phi+ 8k \phi \tilde \delta + \partial_\mu \partial^\mu
\phi-\\ \nonumber &-&
\frac{2}{3}\alpha_{i}\delta_{i}\left(\partial^\rho
\partial^\sigma h_{\rho\sigma}-\partial_\rho \partial^\rho \tilde
h\right)=0.
\end{eqnarray}

Equation (\ref{Lambda}) suggests that there exist preferred values of the parameters
$\alpha_{i}$. Namely, if we  choose these parameters  to be
\begin{equation}\label{alpha}
\alpha_{1}=-\frac{1}{k},\quad \alpha_{2}=\frac{1}{k},
\end{equation}
the values of the cosmological constants on the branes $\Lambda_i$
coincide with the  cosmological constant in the bulk $\Lambda$.
For our choice of the parameters   brane~1 has a positive energy
density, whereas brane~2 has a negative one. We note that one does
not need to worry about the negative sign of the parameter
$\alpha_{1}$: we will see that the model is stable and does not
contain tachyons or ghosts. In fact, our choice of $\alpha_{i}$
does not  introduce any new dimensional parameter and does not
contradict  the naturalness condition.

An interesting observation is that, with  conditions
(\ref{alpha}), equations (\ref{mu-nu}), (\ref{mu-4}), (\ref{4-4})
and (\ref{contracted-44}) possess an additional symmetry under the
transformations
\begin{equation}\label{sym1}
h_{\mu\nu}(x,y)\to
h_{\mu\nu}(x,y)+\sigma\gamma_{\mu\nu}\varphi(x)+\frac{1}{2k^2}\left(\sigma+
\frac{1}{2}\right)\partial_{\mu}\partial_{\nu}\varphi(x),
\end{equation}
\begin{equation}\label{sym2}
\phi(x)\to\phi(x)+\varphi(x),
\end{equation}
which {\it do not} belong to the gauge transformations
(\ref{gaugetrRS}). It is evident that with the help of these
transformations we can impose the condition $\phi(x)\equiv 0$. At
the first glance this symmetry seems to be rather strange. But let
us have a look at equation (\ref{4-4}). It implies  that the
second variation Lagrangian of the theory does not contain the
kinetic term for the field $h_{44}=\phi(x)$ (even in the original
Randall-Sundrum model). It means that the radion field can be
regarded as an auxiliary field (recall the Supersymmetry).
Therefore, there is no contradiction that for some values of
parameters of the model this field can be totally eliminated from
the theory. It should be noted that the symmetry (\ref{sym1}) and
(\ref{sym2}) of the linearized equations of motion can correspond
to some general symmetry of the action (\ref{actionRS}), which is
not evident at first sight (but this is not necessarily so). As we
will see later, if there is matter on the brane, there appears a
scalar field due to the existence of the extra dimension, which
{\it cannot} be identified with the $44$-component of the metric
fluctuations.

It should be noted, that the elimination of the radion in the
Randall-Sundrum model with brane-localized curvature terms was
discussed earlier, for example, in \cite{Luty} and \cite{Padilla}
(it is evident that since there exists some range of parameters
for which the coefficient in front of the kinetic term for the
radion could be either positive or negative, there exist some
values of the parameters for which the radion is absent at all).
But in this paper the equations of motion for the model with
warped background are treated much more thoroughly, a convenient
gauge is used and these equation are solved {\it exactly}.

Thus, we can consider the equations of motion without the radion.
With the help of the regularization
\begin{equation}\label{reg}
\partial_{4}\sigma(\partial^2_{4}\sigma)=\frac{1}{2}\partial_{4}
\left((\partial_{4}\sigma)^2\right)=\frac{1}{2}\partial_{4}k^2=0,
\end{equation}
 we get from  equations (\ref{4-4}) and (\ref{contracted-44})
\begin{equation}\label{tildeh}
\partial_4\tilde h=const\cdot e^{-2\sigma}.
\end{equation} Let us consider Fourier expansion of all terms of
equation (\ref{tildeh}) with respect to coordinate $y$. Since the
term with the derivative $\partial_4$ has no zero mode, this
equation implies that
\begin{equation}
\partial_4(e^{-2\sigma}h)=0,
\end{equation}
where $h=h_{\mu\nu}\eta^{\mu\nu}$. The residual gauge
transformations are sufficient to impose the transverse-traceless
gauge on the field $h_{\mu\nu}$ (see \cite{BKSV})
\begin{eqnarray}\label{TT}
\partial^{\nu}h_{\mu\nu}=0, \\ \nonumber
h=0.
\end{eqnarray}
Thus, the $\mu\nu$-equation takes the form
($\Box=\eta^{\mu\nu}\partial_{\mu}\partial_{\nu}$)
\begin{equation}\label{mu-nu-tt}
\frac{1}{2}\left(e^{-2\sigma}\Box
h_{\mu\nu}+\partial^2_{4}h_{\mu\nu}\right)
-2k^2h_{\mu\nu}+2k\tilde\delta h_{\mu\nu}-\frac{1}{2k}\tilde\delta
e^{-2\sigma}\Box h_{\mu\nu}=0.
\end{equation}
It is not difficult to solve this equation. The zero mode  has the
form
\begin{equation}
h^{0}_{\mu\nu}=\alpha_{\mu\nu}(x)e^{2\sigma},
\end{equation}
whereas the massive modes have the form
\begin{eqnarray}\label{solution}
h^{m}_{\mu\nu}=b^m_{\mu\nu}(x)\Psi^m(y), \quad \Box b^m_{\mu\nu}(x)=m^2 b^m_{\mu\nu}(x),\\
\nonumber
\Psi^{m}(y)=AJ_{2}\left(\frac{m}{k}e^{-\sigma}\right)+
BN_{2}\left(\frac{m}{k}e^{-\sigma}\right),
\end{eqnarray}
 $J_{2}(t)$ and $N_{2}(t)$ being  the Bessel and Neumann
functions.

The term with $\delta$-functions can be taken into account by
imposing the boundary condition
$$AJ_{0}\left(\frac{m}{k}e^{-\sigma}\right)+BN_{0}\left(\frac{m}{k}e^{-\sigma}\right)=0$$
at $y=0$ and $y=R$. The first boundary condition can be satisfied
by an appropriate choice of the coefficients $A$ and $B$:
\begin{equation}\label{besselZ}
\Psi^{m}(y)=N_{m}\left(N_{0}\left(\frac{m}{k}\right)J_{2}\left(\frac{m}{k}e^{-\sigma}\right)-
J_{0}\left(\frac{m}{k}\right)N_{2}\left(\frac{m}{k}e^{-\sigma}\right)\right),
\end{equation}
where $N_{m}$ is the norm of the eigenfunction. The second
boundary condition, at $y = R$, defines the mass spectrum of the
theory and can be rewritten as
\begin{equation}\label{eigenvalues1}
N_{0}\left(\frac{m}{k}\right)J_{0}\left(\frac{m}{k}e^{kR}\right)-
J_{0}\left(\frac{m}{k}\right)N_{0}\left(\frac{m}{k}e^{kR}\right)=0.
\end{equation}
One can see, that it is analogous to the one obtained in
\cite{BKSV}. There exists a theorem about such combinations of
products of Bessel and Neumann functions, which asserts that for
$e^{kR} > 1$ this combination is an even function of $m/k$ and its
zeros are real and simple \cite{BE}. Thus, one does not need to worry
about the stability of the system: there are no tachyons.

The normalized functions $\Psi^{m}(y)$ satisfy the equation (see,
for example, \cite{DHR})
\begin{equation}\label{norm}
\int dy
e^{-2\sigma}\left[1-\frac{1}{k}\delta(y)+\frac{1}{k}\delta(y-R)\right]\Psi^{m}\Psi^{n}=\delta_{mn}.
\end{equation}

Let us calculate the norm of the zero mode eigenfunction
$\Psi^{0}(y)=N_{0}e^{2\sigma}$. Substituting it into (\ref{norm})
one can find that $\Psi^{0}(y)$ can not be normalized, because the
left part of the equation (\ref{norm}) is equal to zero for
arbitrary $N_{0}$. In other words, the second variation Lagrangian
of the theory does not contain the kinetic term for the massless
graviton, i.e. it is absent in this model. One can ask, why do we
consider the model, which does not contain long-range gravity? But
as we will see in the next section, the situation is rather
different, if we place matter on the brane.

\section{Matter on the brane}
Let us suppose that there is  matter on one of the branes (we will specify,
which brane to choose later). Following the DGP-proposal
\cite{DGP}, this matter  induces a brane-localized term
\begin{equation}\label{induced}
S_{ind}=\frac{\Omega^{2}_{ind}}{k16\pi\hat G}\int d^{4}x
\sqrt{\tilde g}\tilde R,
\end{equation}
where $\Omega_{ind}$ is a dimensionless parameter.

Now let us discuss, on which brane we can put the matter (it is
evident that the term (\ref{induced}) on a brane leads to the
redefinition of the corresponding parameter $\alpha_{i}$, for
example, for brane~1 one gets $\alpha_{1}=-\frac{1}{k}\to
\alpha_{1}=\frac{1}{k}(\Omega^{2}_{ind}-1)$). The problem is that
the additional term (\ref{induced}) changes equation
(\ref{mu-nu-tt}) and equation (\ref{eigenvalues1}) for the
eigenvalues, and there may appear tachyonic modes. This situation
was discussed in detail in paper \cite{CharmDuf} (see also
\cite{Shtanov:2003um}). It was shown in these papers that the
gravitational tachyons can be avoided, {\it at least} if (in our
notations used in (\ref{actionRS}))
\begin{equation}\label{tachcond}
\alpha_{1}\ge 0, \alpha_{2}\ge 0
\end{equation}
(it was shown above that tachyons are absent in the case
$\alpha_{1}=-\frac{1}{k}$ and $\alpha_{2}=\frac{1}{k}$ too). So if
$\Omega_{ind}>>1$ and the term (\ref{induced}) arises on the
brane~1 (at y=0), the conditions (\ref{tachcond}) are satisfied
and we do not need to worry about the stability of the model. We
would also like to note that coordinates $x$ are Galilean on
brane~1, so that all the results obtained in coordinates $x$ for
brane~1 are correct from the physical point of view.

One can easily check that the term (\ref{induced}) does not
violate the symmetry (\ref{sym1}), (\ref{sym2}) (because of the
fact that $\sigma|_{y=0}$=0), but only if it arises on brane~1.
Thus, the radion is absent in this model, and we can forget about
the radion as a ghost. On the contrary, the term (\ref{induced})
on  brane~2 (at $y=R$) violates the symmetry (\ref{sym1}),
(\ref{sym2}). Moreover, the existence of the induced terms
(\ref{induced}) on both branes makes the radion to be a ghost (see
\cite{Padilla,CharmDuf}). One can also consider the case
$\sigma=kR-k|y|$ and the existence of matter on  brane~2 only (in
this case coordinates $x$ are Galilean on brane~2). The radion can
be eliminated in this case too (since $\sigma|_{y=R}=0$), but
since $\alpha_{1}=-\frac{1}{k}<0$ and $\alpha_{2}>\frac{1}{k}$
there may appear gravitational tachyons (see
\cite{CharmDuf,Shtanov:2003um}). Thus, the only physically
relevant case, in which tachyons and ghost are absent (and the
symmetry (\ref{sym1}), (\ref{sym2}) is preserved) is when the
matter (and the induced term (\ref{induced})) exists {\it on
brane~1 only}. Brane~2 can be interpreted as a "naked"\ brane,
i.e. a brane without matter on it.

Taking into consideration (\ref{induced}), we get  new equations
of motion (in the case $\phi(x)\equiv 0$)

 1) $\mu\nu$-component
\begin{eqnarray}\label{mu-nu-m}
 & &\frac{1}{2}\left(\partial_\rho \partial^\rho h_{\mu\nu}-
\partial_\mu \partial^\rho
h_{\rho\nu}-\partial_\nu \partial^\rho h_{\rho\mu} +
\partial^2_{4} h_{\mu\nu}\right)-
2k^2h_{\mu\nu}+\frac{1}{2}\partial_\mu
\partial_\nu\tilde h+ \\ \nonumber &+& 2k h_{\mu\nu} \tilde \delta + \frac{1}{2} \gamma_{\mu\nu}\left(\partial^\rho
\partial^\sigma h_{\rho\sigma}-\partial_\rho \partial^\rho \tilde
h - \partial^2_{4} \tilde h-4\partial_4 \sigma\partial_4 \tilde h
\right)-\\ \nonumber &-&
\frac{1}{2k}\tilde\delta\left[\left(\partial_\rho
\partial^\rho h_{\mu\nu}-
\partial_\mu \partial^\rho
h_{\rho\nu}-\partial_\nu \partial^\rho h_{\rho\mu}+\partial_\mu
\partial_\nu\tilde h\right)+\right.\\ \nonumber &+&\left.\gamma_{\mu\nu}\left(\partial^\rho
\partial^\sigma h_{\rho\sigma}-\partial_\rho \partial^\rho \tilde
h\right)\right] + \\ \nonumber &+&
\frac{\Omega^{2}_{ind}}{2k}\delta(y)\left[\left(\partial_\rho
\partial^\rho h_{\mu\nu}-
\partial_\mu \partial^\rho
h_{\rho\nu}-\partial_\nu \partial^\rho h_{\rho\mu}+\partial_\mu
\partial_\nu\tilde h\right)+\right.\\ \nonumber &+&\left.\gamma_{\mu\nu}\left(\partial^\rho
\partial^\sigma h_{\rho\sigma}-\partial_\rho \partial^\rho \tilde
h\right)\right]= -\frac{\hat\kappa}{2}\delta(y)t_{\mu\nu}(x),
\end{eqnarray}

 2) $\mu 4$-component,
\begin{equation}\label{mu-4-m}
\partial_4 ( \partial_\mu \tilde h - \partial^\nu  h_{\mu\nu})= 0,
\end{equation}

 3) $4 4$-component
\begin{equation}\label{4-4-m}
\frac{1}{2}(\partial^\mu \partial^\nu  h_{\mu\nu} - \partial_\mu
\partial^\mu \tilde h ) - \frac{3}{2}\partial_4 \sigma \partial_4 \tilde h=0,
\end{equation}
and
\begin{eqnarray}\label{contracted-44-m}
& &\partial^2_{4} \tilde h + 2\partial_4 \sigma
\partial_4 \tilde h + \frac{2}{3k}\tilde\delta\left(\partial^\rho
\partial^\sigma h_{\rho\sigma}-\partial_\rho \partial^\rho \tilde
h\right)-\\ \nonumber &-&
\frac{2}{3k}\Omega^{2}_{ind}\delta(y)\left(\partial^\rho
\partial^\sigma h_{\rho\sigma}-\partial_\rho \partial^\rho \tilde
h\right)=\frac{\hat\kappa}{3}\delta(y)t_{\mu}^{\mu}(x),
\end{eqnarray}
where $t_{\mu\nu}$ is the energy-momentum tensor of  matter on the
brane, for example, of a static point-like mass. Since
$\sigma|_{y=0}=0$, the existence of matter on brane~1 does not
violate the symmetry (\ref{sym1}), (\ref{sym2}).

To solve these equations, it is convenient to make the following
substitution
\begin{equation}\label{subst}
h_{\mu\nu}(x,y)=b_{\mu\nu}(x,y)+\frac{1}{2k^2}\partial_{\mu}\partial_{\nu}f(x).
\end{equation}
We note  that the second term in  substitution (\ref{subst})
is a pure gauge from the four-dimensional point of view from the
brane.

Let us take equation (\ref{contracted-44-m}). Multiplying it by
$e^{2\sigma}$, using equation (\ref{4-4-m}), taking into account
regularization (\ref{reg}) and the fact that the terms with
derivative have no zero mode, we get
\begin{equation}\label{bt}
\partial^\rho\partial^\sigma b_{\rho\sigma}-\partial_\rho \partial^\rho \tilde
b=-\frac{\hat\kappa k}{2\Omega^{2}_{ind}}\,t^{\mu}_{\mu},
\end{equation}
\begin{equation}\label{bf}
\partial^\rho\partial^\sigma b_{\rho\sigma}-\partial_\rho \partial^\rho
\tilde b+3\partial_\rho \partial^\rho f=0,
\end{equation}
and
\begin{equation}
\partial_{4}\tilde b=0,
\end{equation}
as it was made in Section~2. With the help of the $\mu4$-component
of the equations one can impose the de Donder gauge on the field
$b_{\mu\nu}$ (see also \cite{SV})
\begin{equation}\label{deDond}
\partial^{\nu}\left(b_{\mu\nu}-\frac{1}{2}\gamma_{\mu\nu}\tilde
b\right)=0.
\end{equation}
It follows from  equations (\ref{bt}), (\ref{bf}) and (\ref{deDond})  that
\begin{equation}\label{eqf}
\Box f=\frac{\hat\kappa k}{6\Omega^{2}_{ind}}\,t.
\end{equation}
and
\begin{equation}\label{eqb}
\Box\tilde b=\frac{\hat\kappa k}{\Omega^{2}_{ind}}\,t,
\end{equation}
where $t=\eta^{\mu\nu}t_{\mu\nu}$. We would also like to note that
it is not correct to identify the field $f$ with the radion field
$\phi$, which is the $44$-component of the metric fluctuations.
But it is apparent that this scalar part of $\mu\nu$-component of
the metric fluctuations exists due to the existence of the extra
dimension and appears only if there is some matter on the brane.
It is similar to what happens in the RS2 model, where the radion
field appears, if we place matter on the brane (see \cite{SV2}).

Substituting (\ref{subst}) into (\ref{mu-nu-m}), we  get
\begin{eqnarray}\label{mu-nu-dd}
 & &\frac{1}{2}\left(e^{-2\sigma}\Box
 \left[b_{\mu\nu}-\frac{1}{2}\eta_{\mu\nu}b\right]
+ \partial^2_{4} b_{\mu\nu}\right)- 2k^2b_{\mu\nu}+2k
b_{\mu\nu}\tilde\delta - \\ \nonumber &-&
\frac{1}{2k}\tilde\delta\left(e^{-2\sigma}\Box
\left[b_{\mu\nu}-\frac{1}{2}\eta_{\mu\nu}b\right]\right) +
\frac{\Omega^{2}_{ind}}{2k}\delta(y)\left(e^{-2\sigma}\Box
\left[b_{\mu\nu}-\frac{1}{2}\eta_{\mu\nu}b\right]\right)=\\
\nonumber &=&
-\frac{\hat\kappa}{2}\delta(y)t_{\mu\nu}+\left(1-\frac{\tilde\delta}{k}\right)
\left[\partial_{\mu}\partial_{\nu}f-\eta_{\mu\nu}\Box f\right],
\end{eqnarray}
where $f$ is defined by (\ref{eqf}).

It would be interesting to get the equation for the zero mode of
$h_{\mu\nu}$. As it was noted above, the zero  mode has the
form $h^0_{\mu\nu}=\alpha_{\mu\nu}e^{2\sigma}$. Let us multiply
equation (\ref{mu-nu-dd}) by $e^{2\sigma}$ and integrate it over
coordinate $y$. Using the orthonormality conditions, which is
modified  by the term (\ref{induced}) to be
\begin{equation}\label{norm1} \int dy
e^{-2\sigma}\left[1-\frac{1}{k}\delta(y)+\frac{1}{k}\delta(y-R)+
\frac{\Omega^{2}_{ind}}{k}\delta(y)\right]\Psi^{m}\Psi^{n}=\delta_{mn},
\end{equation}
we  get
\begin{equation}
\Box\left(\alpha_{\mu\nu}-\frac{1}{2}\eta_{\mu\nu}\alpha\right)=-
\frac{\hat\kappa k}{\Omega^{2}_{ind}}\,t_{\mu\nu}.
\end{equation}
An analogous procedure was made in the case of RS1 model in
paper \cite{SV}.

A very interesting thing happens: the massless graviton reappears
in the model. It looks as if the matter "produces"\ the massless
gravity via the induced term (which appears if there is matter on
the brane) for itself. Thus, the gravity on the brane in the zero
mode approximation is defined by
$h^{0}_{\mu\nu}|_{y=0}=\alpha_{\mu\nu}$. The four-dimensional
gravitational constant is defined by the parameter $\Omega_{ind}$
(i.e. by the induced term) instead of the factor $e^{2kR}$ in the
original RS1 model.

\section{Massive modes}

Thus,  we have found a solution for the gravity in the zero mode
approximation. Now let us estimate the effects, which can be produced
by the massive modes. We will not solve equation (\ref{mu-nu-dd}),
as it was made in \cite{SV} for the RS1 model. Let us estimate the
masses of the lowest modes and their wave functions.

An analogue of equation (\ref{mu-nu-tt}) in the presence of the
term (\ref{induced}) has the form
\begin{eqnarray}\label{mu-nu-tt1}
& &\frac{1}{2}\left(e^{-2\sigma}\Box
h_{\mu\nu}+\partial^2_{4}h_{\mu\nu}\right)
-2k^2h_{\mu\nu}+2k\tilde\delta h_{\mu\nu}-\frac{1}{2k}\tilde\delta
e^{-2\sigma}\Box h_{\mu\nu} +\\ \nonumber &+&
\frac{\Omega^{2}_{ind}}{2k}\delta(y)e^{-2\sigma}\Box h_{\mu\nu}=0.
\end{eqnarray}
Following in the footsteps of  Section~2, we  arrive at the
following relations:
\begin{equation}\label{besselZ1}
\Psi^{m}(y)=N_{m}\left(N_{0}\left(\frac{m}{k}e^{kR}\right)J_{2}\left(\frac{m}{k}e^{-\sigma}\right)-
J_{0}\left(\frac{m}{k}e^{kR}\right)N_{2}\left(\frac{m}{k}e^{-\sigma}\right)\right),
\end{equation}
where $N_{m}$ is a normalization constant, and
\begin{eqnarray}\label{eigenvalues2}
&
&N_{0}\left(\frac{m}{k}e^{kR}\right)J_{0}\left(\frac{m}{k}\right)-
J_{0}\left(\frac{m}{k}e^{kR}\right)N_{0}\left(\frac{m}{k}\right)+\\
\nonumber &+&
\Omega^{2}_{ind}\left[N_{0}\left(\frac{m}{k}e^{kR}\right)J_{2}\left(\frac{m}{k}\right)-
J_{0}\left(\frac{m}{k}e^{kR}\right)N_{2}\left(\frac{m}{k}\right)\right]
=0.
\end{eqnarray}
Let us choose $kR$ such that $e^{kR}$ is of the order $10^2\div
10^3$. We can make this assumption, since it is not necessary to
solve the hierarchy problem with the help of the factor $e^{kR}$ -
the four-dimensional Planck mass is defined by $\Omega_{ind}$.
Since $\Omega_{ind}$ is assumed to be much larger than $1$ (we
want to have a small five-dimensional Planck mass), the masses of
the lowest modes are defined by
\begin{equation}\label{eigenvalues3}
J_{0}\left(\frac{m}{k}e^{kR}\right)=0.
\end{equation}
Thus, $m_{low}\sim ke^{-kR}$.

Now let us estimate the normalization constants of the lowest
modes. Using the fact that the Bessel and Neumann functions are
of the order $\sim 1$ for the masses, which are the solutions of
(\ref{eigenvalues3}), from (\ref{norm1}) we get
\begin{equation}\label{norm2}
\frac{1}{N^2_{m}}=\left(\sim
Re^{2kR}\right)-\frac{1}{k}Z^2_{m}(0)+\frac{1}{k}e^{2kR}Z^2_{m}(R)+
\frac{\Omega^{2}_{ind}}{k}Z^2_{m}(0),
\end{equation}
where $\Psi^{m}(y)=N_{m}Z_{m}(y)$. Since $\Omega_{ind}>>e^{2kR}$,
we get
\begin{equation}\label{norm3}
N_{m}\approx \frac{\sqrt{k}}{\Omega_{ind}Z_{m}(0)}.
\end{equation}

Now it is not difficult to calculate the coupling constants of the
massless and massive modes to matter on the brane
($N_{0}=\frac{\sqrt{k}}{\Omega_{ind}}$):
\begin{equation}\label{interaction}
\frac{1}{2}\int_{brane}d^4x \left(\hat\kappa
\frac{\sqrt{k}}{\Omega_{ind}}\alpha_{\mu\nu}(x)t^{\mu\nu}+
\hat\kappa\sum_{m}\frac{\sqrt{k}}{\Omega_{ind}}b^m_{\mu\nu}(x)t^{\mu\nu}\right).
\end{equation}
Identifying the combination
$\frac{\Omega_{ind}}{\hat\kappa\sqrt{k}}$ with the
four-dimensional Planck mass $M_{Pl}$, we get
\begin{equation}\label{interaction1}
\frac{1}{2}\int_{brane}d^4x
\left(\frac{1}{M_{Pl}}\alpha_{\mu\nu}(x)t^{\mu\nu}+
\frac{1}{M_{Pl}}\sum_{m}b^m_{\mu\nu}(x)t^{\mu\nu}\right).
\end{equation}
We see that the coupling constant of the lowest massive modes is
the same, as of the massless graviton.

Now let us discuss the parameters of the model, which can give
some interesting experimental consequences. Taking $M_{*}\sim k$
in the $eV$ range ($M_{*}$ is the five-dimensional Planck mass)
and $kR\sim 5$ we get the size of the extra dimension of the order
of $10^{-5}cm$. The lowest modes have the masses of the order of
$10^{-2}eV$, which correspond to the corrections to  Newton's law
with the strength of the massless graviton at the distances of the
order of $10^{-3}cm$, which is the  micrometer scale. It can be
interesting for experiments on testing gravity at sub-millimeter
scales.

\section{Conclusion}
In this paper a model with brane-localized terms, based on the
Randall-Sundrum background solution for the metric, was considered. The
original Randall-Sundrum scenario was used not for the solution of
hierarchy problem, but as a "constructor"\ of the branes with
tension. Thus, the proper gravitational field of the branes was
taken into account. We showed that this model is free from tachyons
and ghosts and provides some interesting effects, such as the absence of the
radion and "inducing"\  the long-range gravity by matter on the
brane. With an appropriate choice of the parameters, this model can lead to
interesting experimental consequences.

One may ask: why not to choose the original Randall-Sundrum model
with the DGP term ($\sim \Omega^{2}_{ind}$), for example, on
brane~1 only? According to \cite{Padilla,CharmDuf} there are no
gravitational tachyons and ghosts in this case (but the radion
field exists because of the absence of the symmetry (\ref{sym1}),
(\ref{sym2})). But the answer to the question about the radion as
a ghost can be obtained only after {\it the exact} solving  the
equations of motion for the linearized gravity (as it was made,
for example, in \cite{BKSV} for the RS1 model). Nevertheless, if
we retain strong gravity in the bulk (small $M_{*}$, otherwise all
the constructions make no  sense) and $\Omega_{ind}>>e^{2kR}$, the
term (\ref{induced}) on  brane~1 makes the radion to be a ghost
(all the reasonings are the same as in  paper \cite{Luty} for the
$Tev$ brane in the Randall-Sundrum model)! But even if for some
range of parameters the radion is not a ghost (we remind that this
issue can become clear after solving  the corresponding
equations), the possible bounds on the coupling constant of the
radion may create an additional restrictions on the choice of the
parameters $k$ and $R$. At the same time the case of
$\alpha_{2}=\frac{1}{k}$ provides the absence of the radion and
more freedom in our choice of parameters.

We would also like to add that the accuracy in treating  the
equations of motion can provide some interesting results, which
are not evident at the first sight.

It would be interesting to calculate corrections to  Newton's
potential by solving directly equation (\ref{mu-nu-dd}), as it
was made in \cite{SV} for the RS1 model, but this problem deserves
a further investigation.

\bigskip
{\large \bf Acknowledgments}
\medskip \\
The author is grateful to D.~Levkov and especially to
I.P.~Volobuev for valuable discussions. The work was supported by
the grant UR.02.03.002 of the scientific program "Universities of
Russia"\,.

\end{document}